# Single-Input Polarization-Sensitive Optical Coherence Tomography Through a Catheter

GEORGIA L. JONES[1,2], QIAOZHOU XIONG[3], XINYU LIU[4,5], BRETT E. BOUMA[1,2] AND MARTIN VILLIGER [1]

[1] *Wellman Center for Photomedicine, Massachusetts General Hospital, Harvard Medical School, Boston, MA 02114 USA*
[2] *Institute for Medical Engineering and Science, Massachusetts Institute of Technology, Cambridge, MA 02142 USA*
[3] *School of Electrical and Electronic Engineering, Nanyang Technological University, Singapore, 639798, Singapore*
[4] *Singapore Eye Research Institute, Singapore National Eye Centre, Singapore 169856*
[5] *Academic Clinical Program, Duke-NUS Medical School, Singapore 169857*
*\*georgiat@mit.edu*

**Abstract:** Intravascular polarimetry with catheter-based polarization-sensitive optical coherence tomography (PS-OCT) complements the high-resolution structural tomograms of OCT with morphological contrast available through polarimetry. Its clinical translation has been complicated by the need for modification of conventional OCT hardware to enable polarimetric measurements. Here, we present a signal processing method to reconstruct polarization properties of tissue from measurements with a single input polarization state, bypassing the need for modulation or multiplexing of input states. Our method relies on a polarization symmetry intrinsic to round-trip measurements and uses the residual spectral variation of the polarization states incident on the tissue to avoid measurement ambiguities. We demonstrate depth-resolved birefringence and optic axis orientation maps reconstructed from *in-vivo* data of human coronary arteries. We validate our method through comparison with conventional dual-input state measurements and find a mean cumulative retardance error of 13.2deg without observable bias. The 95% limit of agreement between depth-resolved birefringence is $2.80 \cdot 10^{-4}$, which is less than the agreement between two repeat pullbacks of conventional PS-OCT ( $3.14 \cdot 10^{-4}$ ), indicating that the two methods can be used interchangeably. The hardware simplification arising from using a single input state may be decisive in realizing the potential of polarimetric measurements for assessing coronary atherosclerosis in clinical practice.

## 1. Introduction

Intravascular optical coherence tomography (IV-OCT) has become a widely used clinical instrument for high resolution imaging of the coronary arteries during percutaneous coronary intervention (PCI) [1–4]. The technique offers 10 μm depth-resolution with an imaging depth of 1-2 mm [5]. OCT interferometrically measures backscattered light to recover 3D volumetric images of tissue's subsurface microstructure. Although providing high resolution for structural imaging, the OCT scattering signal lacks contrast to differentiate between biologically relevant materials present in coronary atherosclerotic plaques such as collagen, smooth muscle, lipid or macrophages.

To obtain increased compositional contrast, we have been advancing intravascular polarization sensitive optical coherence tomography (IV-PS-OCT) as an extension of conventional IV-OCT [6,7]. PS-OCT not only measures the amplitude of backscattered light, but also its polarization state as a function of the pathlength the light travels into the tissue [8,9]. This enables reconstruction of the polarization properties of the sample, including birefringence and depolarization [10]. Many tissues, especially those with a fibrillar architecture, e.g., collagen, are uniaxially birefringent. Light polarized along and orthogonal to the material's optic axis experiences a difference $\Delta n$ in the refractive index, which causes a general polarization state to change as the light propagates through such a sample. The optic axis furthermore relates to the physical orientation of the fibrillar tissue components. In addition to birefringence, depolarization measures the uniformity of the measured polarization states in a small neighborhood of pixels. Clinical pilot studies of intravascular polarimetry have highlighted the potential of both birefringence and depolarization as a way of identifying important plaque features and differentiating plaque types based on the derived polarization properties [11].

A major barrier to the clinical translation of IV-PS-OCT has been the increased system complexity required to perform polarization-sensitive imaging. Conventional catheter-based PS-OCT systems use two input polarization states as well as polarization diverse detection to fully characterize the polarization properties of the sample. Our previous studies of intravascular polarimetry employed modulation of the sample polarization state between A-lines with an electro-optic modulator. Notably, some clinical intravascular imaging systems have polarization-diverse detection to avoid signal fading due to dynamic polarization change. The ability to perform catheter-based PS-OCT with a single input state would enable such systems to perform polarimetric imaging without any hardware modification but this ability has been unavailable to date.

Benchtop PS-OCT systems frequently use a single circular input state to illuminate the sample [8,12], but this cannot be achieved with a catheter-based system employing a flexible, rotating fiber probe. The probe acts like a rotating waveplate adding to the overall observed round-trip retardance, further precluding the use of reconstructions methods developed for benchtop imaging with a defined circular input state. Yet, the intrinsic reciprocity of OCT's imaging geometry with identical illumination and detection paths ensures that any measured cumulative round-trip retardance is linear, even if system components present in only the illumination or detection paths may break this symmetry [13]. With the prior knowledge that the sample is a linear retarder, it may appear that a single input state is sufficient to determine the unique linear optic axis and amount of retardance that describes the transformation of an input state to the observed output state. However, there is an intrinsic error due to the relative orientation of the input polarization state and the optic axis of the sample [14]. Whenever the observed polarization state corresponds to the input state with reversed helicity, i.e., mirrored by the horizontal plane in Stokes parameter space or the complex conjugate of the Jones vector, the optic axis is undefined, as a suitable retardance to map the input state on this particular output state can be found along any linear optic axis orientation. We previously investigated this phenomenon and termed this output state 'mirror state' [15].

In this paper, we present a reconstruction pipeline for single-input polarization-sensitive optical coherence tomography (SIPS) which leverages the spectral dependence of the polarization state illuminating the sample typically present in fiber-based systems. This residual polarization mode dispersion (PMD) ensures that the sample interacts with a range of polarization states rather than a single one. Our method 1) uses the properties of the mirror state to recover the system symmetry and the spectral dependence directly from typical sample measurements and 2) accurately estimates the cumulative round-trip sample retardance from catheter-based measurements with a single input polarization state. Once the cumulative round-trip signal is recovered, we use established algorithms to derive depth-resolved tissue birefringence and optic axis orientation, and carefully compare the proposed SIPS to conventional dual input PS-OCT (DIPS) using pullbacks of coronary arteries in human patients.

## 2. Methods

For development and validation of SIPS, we used clinical data of coronary arteries, acquired with a conventional dual-input PS-OCT system, described in section A. Each data set was processed independently with both 1) the single-input processing method (SIPS) using only the data corresponding to a single input state, and 2) the conventional dual-input based method (DIPS), as explained in section B. Sections C and D establish our general signal formalism and use of incoherent averaging. We then describe the calibration and reconstruction algorithms for deriving cumulative retardance from catheter-based OCT systems using a single input state. SIPS reconstruction requires knowledge of the system PMD-induced spectral variation present in the illumination and detection paths. This system characterization can be obtained directly from typical measurement data, as described in sections E and F. In the current study, we performed an independent system characterization for each pullback. Once established, it served for SIPS reconstruction of the cumulative round-trip retardance. The reconstruction is formulated as the maximum-likelihood estimation of the most plausible sample retardance, given the observed, spectrally dependent measurements, and is described in detail in section G. After recovering the cumulative retardance with either method, we reconstructed local, depth-resolved tissue birefringence and optic axis orientation (section H). For validation, DIPS was regarded as ground-truth.

### A. PS-OCT System

The IV-PS-OCT system used to collect the clinical data employed commercial intravascular catheters (FastView, Terumo) connected to a custom DIPS OCT system, described previously in detail [11]. In short, the system used a wavelength-swept light source with a repetition rate of 103.6 kHz centered at 1300 nm with a scanning range of 110 nm, resulting in a depth resolution of 9.4 μm in tissue. The system incorporated polarization diverse detection and an electro-optic modulator in the sample arm, alternating the polarization between states located at a relative angle of 90° on the Poincaré sphere. The catheter provided a lateral resolution of ~35 μm, performing 100 rotations per second to acquire frames consisting of 1024 depth-scans, while pulling back at a speed of 20 mm/s.

### B. Dataset Description

This study used an existing clinical dataset of 27 *in-vivo* pullbacks of human coronary arteries. The data was collected at the Erasmus University Medical Center in Rotterdam in 2014-2015 in patients during percutaneous coronary interventions. The study was approved by the Ethics Committee of the Erasmus University Medical Center. All participating patients gave written informed consent. This dataset served previously to investigate the repeatability of DIPS in a clinical setting, by comparing 274 manually identified matching pairs of cross-sections from sequentially acquired pullbacks [16]. Owing to the modulation of the input polarization state, the frames of the DIPS data sets consisting of 1024 depth scans could be split into two sets of SIPS data, containing 512 depth scans each. The limit of agreement between repeat pullbacks measured with DIPS served as the benchmark against which we compared the agreement between SIPS and DIPS.

### C. Mathematical Model of the Single Input PS-OCT Measurement

The transformation of an input polarization state **e** by transmission through the system components and round-trip propagation through the sample can be described using a complex valued 2x2 Jones matrix **G**, producing the output state **t**:

$$\mathbf{t} = \mathbf{G} \cdot \mathbf{e}. \qquad (1)$$

Throughout this manuscript, lowercase bold letters designate vectors, whereas capital bold letters designate matrices. Conventionally, in order to fully characterize a general Jones matrix, at least two linearly independent input states $\mathbf{e}_1$ and $\mathbf{e}_2$ are required, together with the measurements $\mathbf{t}_1$ and $\mathbf{t}_2$, to solve for $\mathbf{G}$. Considering that the only Jones matrix observable in round-trip path is transpose symmetric, $\mathbf{G} = \mathbf{G}^\mathbf{T}$, and furthermore assuming that the only observed polarization effect is retardance, $\mathbf{G}$ is a – possibly scaled – element of the special unitary group of dimension two, SU(2). Assuming further, without loss of generality, that $\mathbf{e} = [1\ 0]^\mathbf{T}$, the Jones matrix can be recovered from a single measured Jones vector $\mathbf{t} = [t_h\ t_v]^\mathbf{T}$

$$\mathbf{G} = \begin{bmatrix} t_h & t_v \\ t_v & -\bar{t}_h \exp(i\,2\,\arg(t_v)) \end{bmatrix}, \qquad (2)$$

where the overbar indicates complex conjugation. This expression is obtained by imposing that the two columns of the Jones matrix must have an inner product that is zero, since $\mathbf{G}^\dagger \cdot \mathbf{G} \propto \mathbf{I}$, where the dagger indicates the complex transpose and $\mathbf{I}$ the identity matrix. Inspecting (2), it is obvious that in the presence of typical noise, small values of $t_v$ cause a large phase uncertainty for the lower right element of $\mathbf{G}$. Indeed, when $\mathbf{t}$ coincides with the mirror state, which is identical to $\mathbf{e}$ for this specific case, the argument of $t_v$ and, hence, $\mathbf{G}$ are undefined.

The presence of single-trip system components in practical fiber-based systems breaks the expected transpose symmetry of the linear retarder [13]. Furthermore, wavelength-dependent polarization effects in the fiber and system components induce PMD, which is managed through the use of spectral binning [17]. Spectral binning splits the recorded spectrum into several, narrower spectral bins. We used 9 overlapping bins spanning each one fifth of the full spectral width. Although the specific scattering amplitude or speckle realization is expected to vary among spectral bins, the underlying polarization effect of the cumulative round-trip Jones matrix of the sample $\mathbf{G}(z)$ is assumed to be independent of the spectral bin and only dependent on pathlength $z$. On the other hand, each spectral bin experiences a distinct system transmission. The pathlength-resolved Jones vectors are modeled as

$$\mathbf{t}(z,p) = \mathbf{C}(p) \cdot \mathbf{Q}^\mathbf{T}(p) \cdot \mathbf{G}(z) \cdot \mathbf{Q}(p) \cdot \mathbf{e}. \qquad (3)$$

$\mathbf{Q}(p)$ represents the Jones matrix of the system's illumination path and depends on the spectral bin $p$, while $\mathbf{C}(p)$ describes the discrepancy of the detection path compared to the illumination path, representing the non-symmetric system effects. We assume that the system matrices $\mathbf{Q}(p)$ and $\mathbf{C}(p)$ are constant across an entire volumetric measurement and only depend on the wavelength, i.e., the spectral bin index $p$. The probe rotation-dependent variation in catheter transmission is assumed independent from the spectral bins and is included within $\mathbf{G}(z)$ and compensated for in later processing as described in Section H. The spectral variation of the system matrices is at the center of our reconstruction method, as it prevents that the polarization states in all spectral bins simultaneously evolve through the mirror state. For convenience, but without loss of generality, we define that at the center spectral bin $p_C$, the symmetric component $\mathbf{Q}(p_C) = \mathbf{I}$, i.e., it equals the identity matrix, by absorbing the symmetric system effect present at the center bin into $\mathbf{J}(z)$.

*D. Incoherent Averaging*

The coherent Jones vector measurements, like any OCT measurements of scattering tissue, are subject to speckle. We transformed Jones vectors to Stokes vectors and used incoherent spatial filtering across A-lines to reduce the detrimental impact of speckle on the measured polarization states, like previously described [17]. We used filtering with a Gaussian window spanning 6 A-lines in width. For DIPS processing, the same filtering was applied independently to the Stokes vectors of the two input states. Transforming a Jones to a Stokes vector eliminates the global phase of the Jones vector, without consequence for the present polarization analysis, as shown below. If we have a Jones vector **t**, with complex components $t_h$ and $t_v$ representing the electric fields, the transformation to the corresponding Stokes vector **s** is

$$\mathbf{t} = \begin{bmatrix} t_h \\ t_v \end{bmatrix} \leftrightarrow \mathbf{s} = \begin{bmatrix} I \\ Q \\ U \\ V \end{bmatrix} = \begin{bmatrix} |t_h|^2 + |t_v|^2 \\ |t_h|^2 - |t_v|^2 \\ 2\Re\{t_h \bar{t}_v\} \\ -2\Im\{t_h \bar{t}_v\} \end{bmatrix}. \qquad (4)$$

Here, $\Re$ denotes the real part of the enclosed expression and $\Im$ denotes the imaginary part. Importantly, the polarized components of a Stokes vector can be converted back to a coherent Jones vector, leaving simply an ambiguous global phase:

$$t_h = \sqrt{\frac{I+Q}{2}}, \ t_v = \frac{U+iV}{\sqrt{2}\sqrt{I+Q}}. \qquad (5)$$

Throughout this manuscript, we use SU(2) notation whenever possible, but employ the isomorphism between SU(2) and the special orthogonal group SO(3), which describes the identical polarization transformations on fully polarized Stokes vectors. The averaged Stokes vectors were also used to derive the degree of polarization of the incident light at every pixel, although we use depolarization, defined as one minus the degree of polarization for visualization. Depolarization can be used as a signal validity metric, to mask areas where birefringence measurements are unreliable, but also highlights specific tissue structures that randomize the polarization states of the scattered light.

*E. System Characterization: Asymmetric Component*

The asymmetric portion of the system effect, $\mathbf{C}(p)$ is estimated using the mirror state phenomenon [15]. For a symmetric system, the mirror state corresponds to the input state with reversed helicity, i.e., its complex conjugate Jones vector. A possible asymmetric component simply further transforms the symmetric into the observed mirror state $\mathbf{m}(p)$:

$$\mathbf{m}(p) = \mathbf{C}(p) \cdot \bar{\mathbf{e}}. \qquad (6)$$

The pathlength-dependent polarization state evolution due to round-trip propagation into a homogeneously retarding sample must pass through the mirror state [15]. As a result, the mirror state is the most commonly measured polarization state if a sufficient amount of heterogenous sample data is analyzed. Hence, we computed histograms of the observed Stokes vectors by defining bins of uniform area on the surface of the Poincaré sphere for each pullback. The histograms were computed using the filtered Stokes vectors of pixels with low depolarization ($< 0.3$) from 20 cross sections for each pullback. Experimental histograms may exhibit multiple local maxima, and we furthermore modeled the spectral dependence of the asymmetric system component C(p) as an element subject to first order PMD [18]:

$$\mathbf{C}(p) = e^{i\mathbf{A}\cdot(p-p_C)} \cdot \mathbf{C}_0, \qquad (7)$$

where $p_C$ is the center spectral bin. This expression corresponds to the product of a general retarder $\mathbf{C}_0$ and a first-order wavelength-dependent retarder, where $\mathbf{A}$ is a matrix with Hermitian symmetry and a Frobenius norm proportional to the retardance between two neighboring spectral bins. Together with $\mathbf{e} = [1\ 0]^\mathbf{T}$, this offered a model for $\mathbf{m}(p)$ that could be fit to the potentially multiple local maxima in the recovered histograms using random sample consensus (RANSAC) [19] for robust estimation of $\mathbf{C}(p)$. The process of mirror state estimation is graphically represented by Fig. 1.

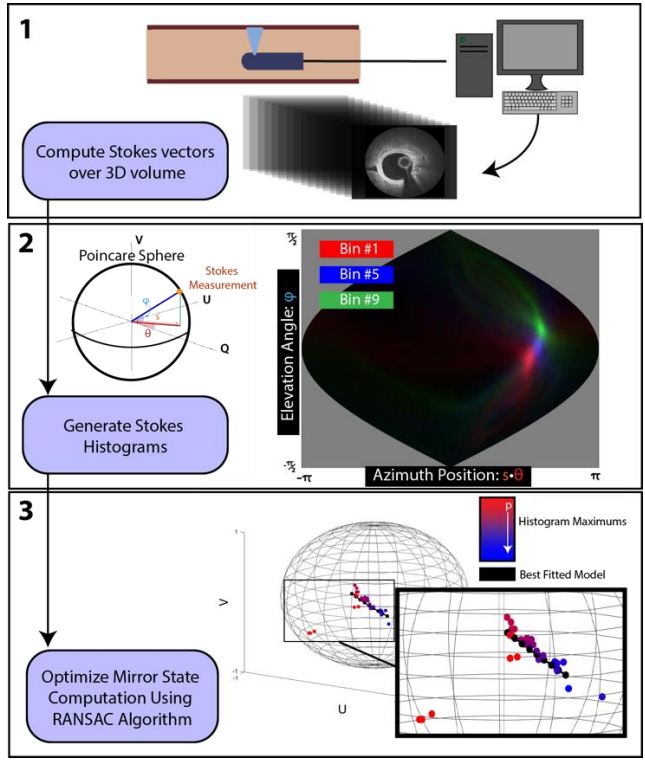

Figure 1: Description of the process of computing the system symmetrization matrix C using the mirror state. Panel 1 indicates the extraction of Stokes vectors over a large collection of data. Panel 2 displays the Stokes histograms of the Poincaré surface for the first (red), central (blue) and final (green) bin. The y-axis encodes the elevation angle $\varphi$ on the Poincaré sphere and the x-axis encodes the azimuthal arc length $s \cdot \theta$, where $s = cos(\varphi)$. Panel 3 indicates the RANSAC-determined best fit of the mirror state corresponding to C(p) as a function of spectral bin. The dots on the Poincare sphere each indicate a mirror state identified in the histograms shown in Panel 2, with the color indicating the spectral bin. The black dotted line indicates the best fit for the system mirror state given the C(p) acts as first order PMD.

## F. System Compensation: Symmetric Component

After computation of and compensation for the asymmetric component $\mathbf{C}(p)$, the remaining system effect can be viewed as a spectral-bin-dependent general retarder, denoted by $\mathbf{Q}(p)$ and acting transpose-symmetrically in the illumination and detection paths

$$\tilde{\mathbf{t}}(z,p) = \mathbf{C}^{-1}(p) \cdot \mathbf{t}(z) = \mathbf{Q}^{\mathbf{T}}(p) \cdot \mathbf{G}(z) \cdot \mathbf{Q}(p) \cdot \mathbf{e}. \quad (8)$$

The tilde indicates the Jones vector after correction for the asymmetric component $\mathbf{C}(p)$. Following (2), we use the same data points that already served to retrieve $\mathbf{C}(p)$, to recover their associated transpose symmetric matrices.

$$\mathbf{G}_{QQ}(z,p) = \mathbf{Q}^{\mathbf{T}}(p) \cdot \mathbf{G}(z) \cdot \mathbf{Q}(p). \quad (9)$$

Since we defined $\mathbf{Q}(p_c)$ as the identity matrix, we are looking for the $\mathbf{Q}(p)$ of the other spectral bins that align the observed sample matrices, by averaging over the $n$ measured, corrected Stokes vectors of this set of data points:

$$\min_{\mathbf{Q}(p)} \sum_n \left| \mathbf{Q}(p)^{-\mathbf{T}} \cdot \mathbf{G}_{QQ}(z,p) \cdot \mathbf{Q}(p)^{-1} - \mathbf{G}_{QQ}(z,p_C) \right|^2 \quad (10)$$

where **-T** indicates the combined inverse and transpose operation. To avoid challenges with the global phase of $\mathbf{G}_{QQ}$, we solve the corresponding minimization problem in Stokes formalism using fminsearch in MATLAB, by parameterizing $\mathbf{Q}(p)$ as a general linear retarder. Although $\mathbf{G}_{QQ}$ is inaccurate when the observed polarization state aligns with the mirror state, the resulting error is unbiased and averaging over many data points offers robust estimation of $\mathbf{Q}(p)$. The minimization in (10) results in a term that is linear in $\mathbf{G}_{QQ}$, allowing to compute the summation over the data points before multiplication with $\mathbf{Q}(p)$, making this optimization process very efficient. This process is thoroughly described in Supplementary Material (**S.1**).

Although this enables robust estimation of $\mathbf{C}(p)$ and $\mathbf{Q}(p)$, there remains some ambiguity with respect to $\mathbf{C}(p)$, as any retarder $\mathbf{E}$ of which $\mathbf{e}$ is an eigenstate offers an alternative symmetrization matrix $\mathbf{C'}(p)$ and corresponding alignment matrix $\mathbf{Q'}(p)$.

$$\underbrace{\mathbf{C} \cdot \mathbf{E}^{-\mathbf{T}}}_{\mathbf{C'}} \cdot \underbrace{\mathbf{E}^{\mathbf{T}} \cdot \mathbf{Q}^{\mathbf{T}}}_{\mathbf{Q'}^{\mathbf{T}}} \cdot \mathbf{G} \cdot \underbrace{\mathbf{Q} \cdot \mathbf{E}}_{\mathbf{Q'}} \cdot \underbrace{\mathbf{E}^{-1} \cdot \mathbf{e}}_{\mathbf{e'}}$$
$$= \mathbf{C'} \cdot \mathbf{Q'^T} \cdot \mathbf{G} \cdot \mathbf{Q'} \cdot \mathbf{e'}. \quad (11)$$

If $\mathbf{e}$ is an eigenstate of the retarder $\mathbf{E}$, the altered input state $\mathbf{e'}$ is different simply by a phase offset ($\mathbf{e'} = e^{ip} \mathbf{e}$), where the phase $p$ is any real number, resulting in an alternative set of system matrices associated with the same input state.

### G. Determine Sample Roundtrip *Retardance with Singular Value Decomposition*

At this point, we have corrected the measurements for the asymmetric system components and know how the symmetric system transmission varies between spectral bins. Instead of constructing the transpose-symmetric matrix $\mathbf{G}_{QQ}$ independently for each bin, which is inaccurate when a polarization state close to the mirror state is measured, we use the now established knowledge of $\mathbf{Q}(p)$ to directly estimate a unitary $\mathbf{J}(z)$ across all bins simultaneously:

$$\varepsilon = \sum_p \varepsilon_p = \sum_p \left\| \tilde{\mathbf{t}}_p - c_p \, \hat{\mathbf{t}}_p \right\|^2$$

$$= \sum_p \left\| \tilde{\mathbf{t}}_p - c_p \, \mathbf{Q}_p^T \cdot \mathbf{J} \cdot \mathbf{Q}_p \cdot \mathbf{e} \right\|^2, \quad (12)$$

where $\varepsilon$ is the error across spectral bins, $\hat{\mathbf{t}}_p$ is the estimated measurement given a certain "proposed", unitary Jones matrix $\mathbf{J}$. We also omitted the variable $z$ and simplified the notation by writing $p$ as subscript. By construction $\hat{\mathbf{t}}_p$ has unit norm. A complex-valued coefficient $c_p$ is added to manage the phase and amplitude between the measured and estimated output states. The estimated output Jones vector, $\hat{\mathbf{t}}_p$ is the product of the proposed cumulative sample matrix $\mathbf{J}$, the matrices $\mathbf{Q}_p$ and the input state $\mathbf{e}$. The "best" guess for $\mathbf{J}$ reduces the total error across all bins. Noise in complex-valued OCT tomograms is commonly modeled as complex-valued, additive, and normally distributed. Under these assumptions, (12) can be identified as the maximum-likelihood estimation of $\mathbf{J}$.

*1. Solving for $c_p$*

We define $c_p$ as the complex value that minimizes the error between an estimated and measured Jones vector

$$\min_{c_p} \varepsilon_p = \min_{c_p} \left\| \tilde{\mathbf{t}}_p - c_p \, \hat{\mathbf{t}}_p \right\|^2$$
$$= \min_{c_p} \left( \left\| \tilde{\mathbf{t}}_p \right\|_2^2 + c_p \bar{c}_p \left\| \hat{\mathbf{t}}_p \right\|_2^2 - \tilde{\mathbf{t}}_p^\dagger \cdot \hat{\mathbf{t}}_p c_p - \hat{\mathbf{t}}_p^\dagger \cdot \tilde{\mathbf{t}}_p \bar{c}_p \right). \quad (13)$$

Solving for $c_p$ we find that it scales the modeled unit vector to match the measured one in both amplitude and phase

$$c_p = \hat{\mathbf{t}}_p^\dagger \cdot \tilde{\mathbf{t}}_p. \quad (14)$$

*2. Simplifying Expression*

Including the $c_p$ of (14) in (13), we can simplify the minimization of the total error to

$$\arg\min_{\mathbf{J}} \varepsilon = \arg\min_{\mathbf{J}} \sum_p \left( \left\| \tilde{\mathbf{t}}_p \right\|_2^2 - \left\| \hat{\mathbf{t}}_p^\dagger \cdot \tilde{\mathbf{t}}_p \right\|_2^2 \right)$$
$$= \arg\max_{\mathbf{J}} \sum_p \left\| \hat{\mathbf{t}}_p^\dagger \cdot \tilde{\mathbf{t}}_p \right\|_2^2. \quad (15)$$

Representing the scalar product of two vectors as a trace, the cyclic property of traces can be used to reformat the expression such that the Jones matrix appears on the right-hand side:

$$\tilde{\mathbf{t}}_p^\dagger \cdot \hat{\mathbf{t}}_p = \mathrm{tr}(\tilde{\mathbf{t}}_p^\dagger \cdot \hat{\mathbf{t}}_p) = \mathrm{tr}(\tilde{\mathbf{t}}_p^\dagger \cdot \mathbf{Q}_p^T \cdot \mathbf{J} \cdot \mathbf{Q}_p \cdot \mathbf{e}) = \mathrm{tr}(\mathbf{Q}_p \cdot \mathbf{e} \cdot \tilde{\mathbf{t}}_p^\dagger \cdot \mathbf{Q}_p^T \cdot \mathbf{J}) = \mathrm{tr}(\mathbf{M}_p \cdot \mathbf{J}) = \vec{\mathbf{M}}_p^\dagger \cdot \vec{\mathbf{J}}.$$
$$(16)$$

We define a measurement matrix $\mathbf{M}_p = \overline{\mathbf{Q}}_p \cdot \tilde{\mathbf{t}}_p \cdot \mathbf{e}^\dagger \cdot \mathbf{Q}_p^\dagger$, which in addition to the Jones matrix $\mathbf{J}$ is column-wise vectorized, indicated by the arrow symbol. The maximization of this term can be then expressed as

$$\operatorname*{argmax}_{\mathbf{J}} \left\| \vec{\mathbf{t}}_p^{\dagger} \cdot \tilde{\mathbf{t}}_p \right\|_2^2 = \operatorname*{argmax}_{\mathbf{J}} \vec{\mathbf{J}}^{\dagger} \cdot \underbrace{\left( \sum_p \vec{\mathbf{M}}_p \cdot \vec{\mathbf{M}}_p^{\dagger} \right)}_{\mathbf{M}} \cdot \vec{\mathbf{J}}$$

$$= \operatorname*{argmax}_{\mathbf{J}} \vec{\mathbf{J}}^{\dagger} \cdot \mathbf{M} \cdot \vec{\mathbf{J}}. \qquad (17)$$

The bracketed sum gives a bin-independent full-rank 4x4 matrix **M** consisting only of the measurements $\tilde{\mathbf{t}}_p$ and the known quantities **e** and $\mathbf{Q}_p$.

*3. Reparameterizing* **J**

Maximizing the above expression in closed form remains difficult due to the complex-valued vector $\vec{\mathbf{J}}$. Instead, we recall that **J** is a transpose symmetric linear retarder and can be expressed as

$$\mathbf{J} = \begin{bmatrix} \alpha - i\beta & -i\gamma \\ -i\gamma & \alpha + i\beta \end{bmatrix}, \qquad (18)$$

where

$$\alpha = \cos\left(\frac{\varphi}{2}\right) \qquad (19)$$

$$\beta = \sin\left(\frac{\varphi}{2}\right) \cos(2\theta) \qquad (20)$$

$$\gamma = \sin\left(\frac{\varphi}{2}\right) \sin(2\theta) \qquad (21)$$

$$\alpha^2 + \beta^2 + \gamma^2 = 1, \qquad (22)$$

with $\varphi$ encoding the retardance and $\theta$ the optic axis orientation. We can now define the real-valued vector $\mathbf{j} = [\alpha, \beta, \gamma]^{\mathrm{T}}$ which has unit length to express the vectorized sample Jones matrix **J** as

$$\vec{\mathbf{J}} = \sqrt{2} \cdot \mathbf{P} \cdot \begin{bmatrix} \alpha \\ \beta \\ \gamma \end{bmatrix} = \sqrt{2} \cdot \mathbf{P} \cdot \mathbf{j}, \qquad (23)$$

where

$$\mathbf{P} = \frac{1}{\sqrt{2}} \begin{bmatrix} 1 & -i & 0 \\ 0 & 0 & -i \\ 0 & 0 & -i \\ 1 & i & 0 \end{bmatrix}. \qquad (24)$$

Note that $\mathbf{P}^{\dagger} \cdot \mathbf{P} = \mathbf{I}$.

*4. Error Minimization*

With the reparameterization of **J** through **j**, the maximization can be rewritten as

$$\max_{\mathbf{j}} \mathbf{j}^\dagger \cdot \mathbf{P}^\dagger \cdot 2\mathbf{M} \cdot \mathbf{P} \cdot \mathbf{j} = \max_{\mathbf{j}} \mathbf{j}^\dagger \cdot \Re\{\mathbf{H}\} \cdot \mathbf{j}. \qquad (25)$$

$\mathbf{H} = 2\,\mathbf{P}^\dagger\!\cdot\!\mathbf{M}\!\cdot\!\mathbf{P}$, by construction, is a Hermitian, positive definite, 3x3 matrix. Because $\mathbf{j}$ is real-valued, the imaginary part of $\mathbf{H}$ does not contribute to the inner product of $\mathbf{j}$. Furthermore, because $\mathbf{j}$ has unit length, the solution to this constrained maximization problem is the singular vector corresponding to the largest singular value of the real part ($\Re$) of the matrix $\mathbf{H}$. The more formal derivation of the solution to this optimization problem is detailed in the Supplementary Material (**S.2**).

### H. Depth-resolved birefringence and optic axis orientation

Once the ideal cumulative vectorized Jones matrix $\mathbf{j}$ is determined, the local, depth-resolved polarization properties of the sample, i.e., birefringence and optic axis orientation can be derived in the same manner as in conventional DIPS [13,17]. These cumulative matrices have undergone compensation for PMD in the OCT system, but still need correction for the transmission through the catheter to obtain meaningful optic axis measurements. In short, we detect the round-trip signal from the inner sheath to model and compensate the catheter transmission, and use the retardance across the sheath as absolute optic axis reference. Then we iteratively recover the local linear retarder of each tissue layer with a depolarization < 0.2.

Starting from the luminal surface, we determine the cumulative round-trip linear retardance to a given tissue depth and compensate it with the inverse of the single-pass transmission to a one-pixel shallower depth, both for the illumination and the detection. Taking the square root of the roundtrip transmission through this isolated layer provides the depth-resolved single-pass retardance and optic axis orientation of this layer, and allows updating the single pass transmission through all preceding layers to then isolate the next deeper tissue layer. Therefore, iterative reconstruction is only necessary for the depth-resolved optic axis orientation. Scalar depth-resolved birefringence and depolarization were obtained non-iteratively.

### 3. Results

SIPS relies on the inherent PMD of the system, which transforms the single input polarization state into a range of polarization states which interact with the sample. By 1) estimating the polarization mode dispersion of the system and 2) intelligently managing the wavelength-dependent sample information, the cumulative retardance matrix of the sample can be extracted. For the system compensation, we developed a framework to estimate both the asymmetric and symmetric effect of the system directly from heterogenous data, without requiring specific calibration measurements. After compensating for the system effect, maximum likelihood estimation is used to accurately manage the orientation-dependent error of polarization measurements and derive the most likely sample retardance, as described in detail in the Methods section. Once the cumulative retardance is recovered, it can be processed like DIPS data to retrieve local, depth-resolved birefringence and optic axis orientation. Presently, we will qualitatively and quantitatively compare our presented SIPS algorithm to the conventional DIPS reconstruction, for both cumulative round-trip and local retardance properties (see graphical overview in Visualization 1).

Running in MATLAB on the CPU, SIPS processing takes approximately 10% longer than for conventional DIPS (5.4 seconds versus 4.9 seconds per cross-section). We are confident that optimizing the algorithmic implementation and the use of GPU or FPGA will be able to substantially increase processing speed.

*A. Qualitative Comparison of SIPS and DIPS*

Fig. 2 displays the local birefringence, depolarization, and depth-resolved optic axis orientation images derived by SIPS (a-c) and conventional DIPS (d-f) for a representative cross-section. Fibrillar components tend to have higher birefringence value compared to cellular components. One particularly fibrillar area in the coronary artery is the tunica media, a circumferential layer of smooth muscle cells approximately 200μm in depth.

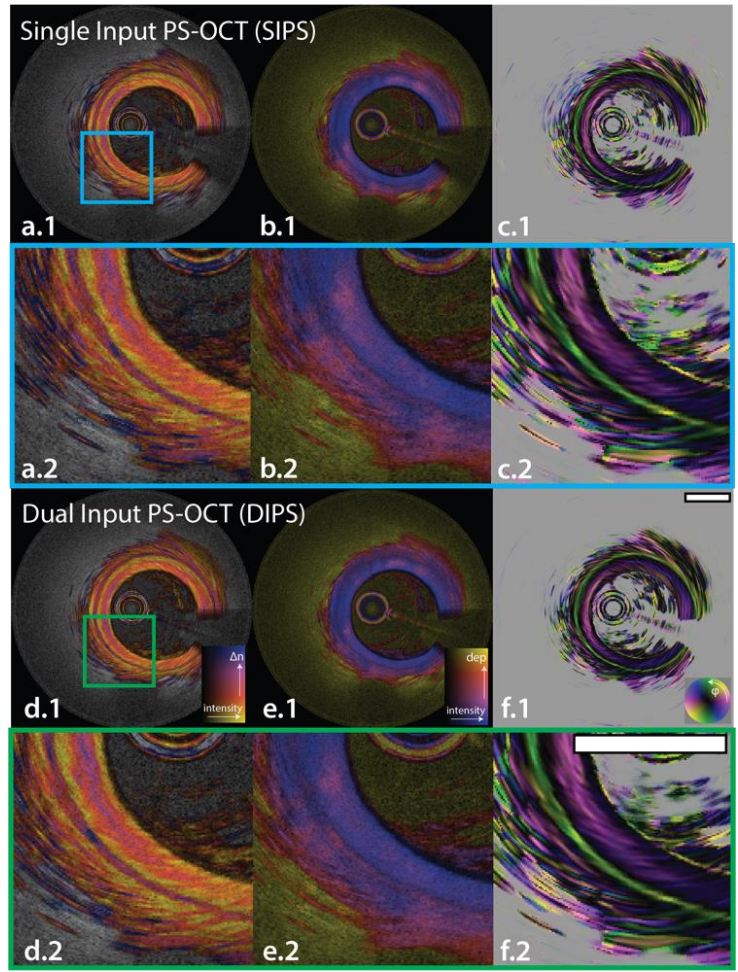

Figure 2: Qualitative comparison of the polarization property images of the presented maximum-likelihood based SIPS (left) and conventional DIPS (right). Image groupings a&d present birefringence ($\Delta n$), b&e depolarization (Dep), and c&f the local optic axis orientation images ($\varphi$), of SIPS and DIPS, respectively. Birefringence is scaled from 0 to $2.2 \cdot 10^{-3}$. Depolarization is scaled between 0.5 and 1. Scale bars indicate 1mm.

In Fig. 2, this layer can clearly be visualized in both the SIPS and DIPS images (a.2. and d.2.) as a layer of increased birefringence. In addition, the corresponding optic axis orientation shows a consistent, tangential orientation in the media region, which aligns with the expected fiber morphology. The DIPS and SIPS methods show clear qualitative agreement in all polarization parameters, including depth-resolved birefringence and optic axis orientation. The next sections will quantify this agreement both in direct pixel comparisons and in the context of measurement repeatability.

*B. Quantitative Comparison of Cumulative Retardance Matrices*

Firstly, we compare the cumulative round-trip retardance matrices between SIPS and DIPS on a pixel-by-pixel basis. Cumulative retardance matrices were chosen as all proceeding steps (such as local property extraction) are identical between the SIPS and DIPS pipeline.

Comparing cumulative matrices allows for the direct comparison of the two methods without having to contend with existing issues further down in the processing, for example with depth-resolved optic axis reconstruction through a catheter. The quantitative difference between the cumulative retardance matrices derived by SIPS versus DIPS was determined using the Riemannian distance between pairs of SO(3) matrices [20].

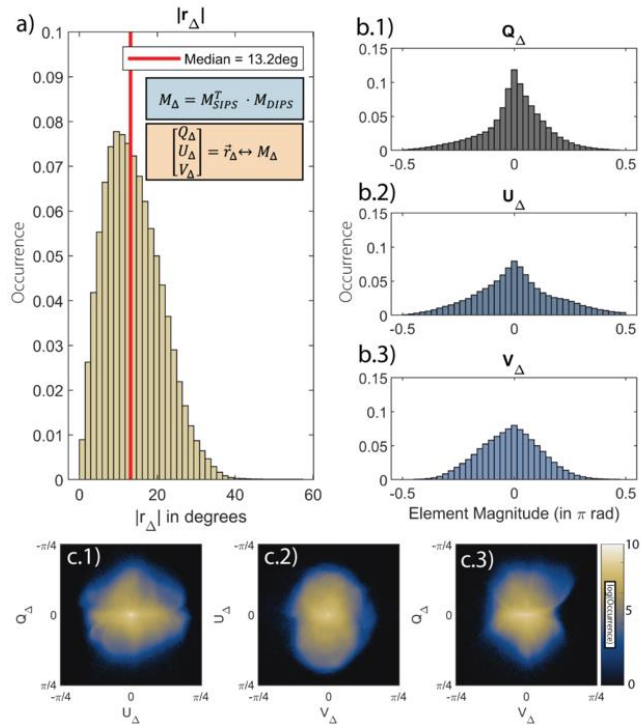

Figure 3: Quantitative comparison of cumulative retardance matrix MSIPS and MDIPS in SO(3) after correction with estimated pullback-dependent Q-oriented linear retarder **E**. a) The distribution of the error vector magnitude, with a median length of ~13.16deg (0.26rad). This is the median difference in cumulative retardance between the conventional DIPS and the proposed SIPS method. b) Histograms of the individual components of the error vector of the Q (b.1), U (b.2.) and V (b.3.) axis. The error vector magnitude is centered around [Q,U,V] = [0.001,0.010,-0.018] radians with standard deviations of 0.155rad, 0.186rad and 0.141rad for the Q, U and V axis, respectively. c) Histograms of the logarithm of the error vector r$\Delta$ magnitude along c.1 the QU plane, c.2 the UV plane and c.3 the QV plane.

This distance corresponds to the length of rotation vector **$r_\Delta$**, which describes the rotation required to align a SIPS cumulative matrix to its corresponding DIPS cumulative matrix in SO(3). If the two matrices are identical **$r_\Delta$** is a zero vector. 10 cross sectional images were randomly selected from each of the 27 patient pullbacks. From each of these images, 5000 data points were randomly selected out of all pixels with sufficiently low depolarization (<0.2) and outside of the catheter sheath, amounting to approximately 13.5 million data points. These data points served to characterize the difference between SIPS and DIPS. The depolarization threshold was chosen to ensure high accuracy polarization measurements in analyzed data as well as maintain consistency with previous results on repeatability [16]. Due to the inherent ambiguity in the asymmetry compensation, the SIPS-derived cumulative retardance matrix may differ from the DIPS-derived matrix by an unknown Q-oriented linear retarder **E**, acting both on its input and output sides. Note that **E** has become restricted to a Q-oriented linear retarder due to the assumption that the input state **e** is horizontally polarized. Given the constraint that **e** must be an eigenstate of the retarder **E**, only Q-oriented retarders maintain this assumption.

**E** was estimated using 2000 pairs of SIPS and DIPS data points for each pullback and applied consistently across all remaining data points of the pullback (500,000) prior to the pixel-by-pixel comparison. This ambiguity does not affect any of the proceeding local property computations and was only corrected for this pixel-by-pixel analysis of cumulative retardance.

Fig. 3a displays the histogram of error vector magnitude over all analyzed points, resulting in a median cumulative retardance error of 13.2 deg. Fig. 3b shows the distribution of individual error vector components along the three axes (Q, U, V) of the Poincare sphere and indicates that the error is centered around the origin. Fig. 3c presents the 2D histograms of $r_\Delta$ for each orthogonal plane of the Poincare sphere, summed over the remaining variable.

*C. Quantitative Comparison of Local Properties*

To contextualize the relative error between SIPS and DIPS, we compared it to the inter-pullback repeatability error of DIPS. A previous study utilized repeat pullbacks in the same coronary artery of human patients to get an estimate of the inter-pullback error of birefringence.

The comparison was done using Bland-Altman analysis, which evaluates the agreement between two measurements by plotting the absolute difference in measurement values versus the average measurement value, to reveal any constant or value-dependent bias. The difference furthermore serves to define the limit of agreement, expressing a bound on the error between the two measurements within which a certain percentage, e.g., 95%, of the measurements fall. Fig. 4a & 4b present two Bland-Altman histograms, **a,** repeating the previous investigation of comparing the *inter-pullback* error of conventional DIPS as a reference and **b**, comparing the *inter-method* error of SIPS versus DIPS.

Fig. 4c presents the normalized sum over all birefringence errors (denoted the error histogram) of three *inter-pullback* analyses and one *inter-method* analysis (SIPS vs DIPS). As SIPS processing uses only half of the available depth scans (see Section 2.B), the inter-pullback error histogram of SIPS versus SIPS (Fig. 4c, green line) should be comparable to the inter-pullback error histogram of DIPS versus DIPS computed with only half of the available modulated A-lines (Fig. 4c, yellow). Fig. 4c shows that the inter-pullback error histogram of SIPS versus SIPS has a smaller limit of agreement than the sub-sampled DIPS versus DIPS error histogram, suggesting that SIPS is more robust to inter-pullback variation than conventional DIPS. This data was obtained by using low depolarization (<0.2) points outside the catheter-sheath from registered cross sections from 27 patients. The local properties were averaged over registered circular regions of interests, 300 μm in diameter. We found 95% limits of agreement of $3.14 \cdot 10^{-4}$ for the *inter-pullback* comparison and $2.80 \cdot 10^{-4}$ for the *inter-method* error. This highlights that the error between the local birefringence values of SIPS and DIPS is not greater than the error between two pullbacks of conventional DIPS, and that SIPS and DIPS can be used interchangeably.

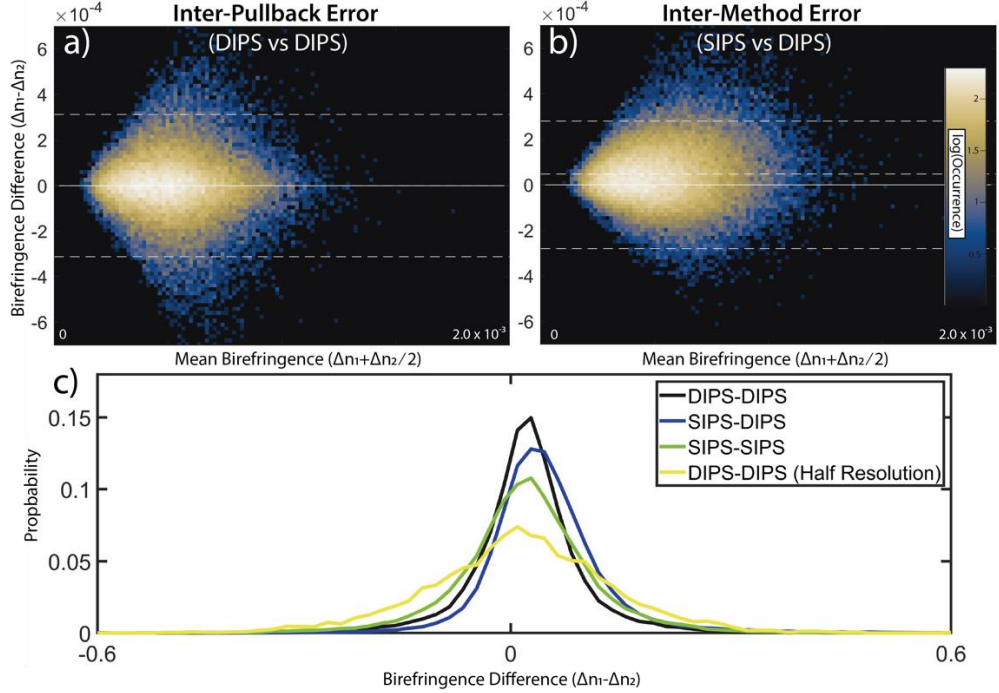

Figure 4: Bland Altman plots comparing 1) Inter-pullback DIPS versus DIPS, 2) inter-method SIPS versus DIPS. The logarithm of the histogram is visualized on both Bland-Altman plots. The horizontal axis encodes the mean value between the two methods being compared (two repeat pullbacks in **a)** and two methods in **b)**. The vertical axis encodes the difference between the two measured values (range is $[-7 \cdot 10^{-4}, 7 \cdot 10^{-4}]$) while the horizonal axis encodes the mean of the two values (range is $[0, 2.0 \cdot 10^{-3}]$)). The bias in **a)** is $-3.43 \cdot 10^{-7}$, and the limit of agreement (LOA) $3.14 \cdot 10^{-4}$. In **b)** the bias is $4.74 \cdot 10^{-5}$ and the LOA $2.80 \cdot 10^{-4}$. **c)** Histograms of birefringence difference for both inter-method (SIPS-DIPS in blue) and inter-pullback (DIPS-DIPS in black, SIPS-SIPS in green and DIPS-DIPS at half resolution in yellow) analysis.

## 4. Discussion

Polarization-sensitive optical coherence tomography (PS-OCT) has significant potential as both a clinical and research tool. The modality combines the micron-level resolution of conventional structural optical coherence tomography with insightful contrast of polarimetry. In interventional cardiology, PS-OCT may facilitate the identification of important plaque components and inform treatment decisions [7]. Both benchtop and catheter-based PS-OCT have shown promise as diagnostic or guidance tool in other fields including ophthalmology, oncology, pulmonology and neurology [21–27]. As the birefringence values computed with SIPS are comparable to those of conventional DIPS, we expect similar contrast in applications where PS-OCT has previously shown promise.

In an effort to realize the clinical potential of PS-OCT, we have developed a signal-processing pipeline that enables robust reconstruction of polarization properties with only a single input state, enabling easier clinical translation. The only system requirements are polarization diverse detection and sufficient system PMD. This approach offers the ability to look back on previously acquired datasets meeting these system requirements and re-evaluate studies by considering additional polarimetric information. While coronary artery imaging has been the clinical focus of this preliminary work, the presented signal processing pipeline remains applicable to all OCT systems, including benchtop configuration, with sufficient PMD and polarization diverse detection.

Qualitatively, human coronary images reconstructed with conventional DIPS and SIPS display significant similarity in all evaluated polarization properties. The difference in local birefringence between SIPS and DIPS was almost identical to that between two pullbacks of the same coronary artery processed with DIPS. This indicates that the error of computing a cross section with SIPS rather than DIPS is no greater than the inherent error expected with taking a repeat measurement.

In previous work, we utilized the mirror state to obtain depth-resolved birefringence measurements [15], but relied on conventional DIPS processing to obtain the system correction matrices and used a sub-optimal weighting scheme for combining spectral bin information. Our current methodology for SIPS overcomes these limitations and demonstrates robust recovery of the pathlength-resolved cumulative round-trip retardance, which then serves for computation of the local, depth-resolved tissue birefringence and optic axis orientation.

More explicitly than DIPS, SIPS relies on the assumption that the sample is a pure retarder. Reconstruction of sample diattenuation would require measurement of a more general, non-unitary Jones matrix, which is currently only possible with PS-OCT systems that depth-multiplex two input states [28,29], at the cost of increased system complexity. Resourceful use of polarization maintaining fiber affords multiplexing of input states combined with polarization-diverse detection [9], but is incompatible with existing clinical systems. Additionally, there have been studies that use single-input polarization state measurements to depth-resolve birefringence and optic axis orientation in benchtop systems [30], however without taking into account the dependence of the measurement reliability on the alignment between the observed polarization state and the mirror state.

It is important to note that although the mathematical framework describes the input state **e** as horizontally polarized, this is an ambiguity that is contained within the asymmetry matrix **C**. Even if the true input state were different, this offset is becoming integral part of **Q**, and is compensated by the determined **C**. This allows for the input state to take any form, as long as the system compensation matrices have been computed from measurements taken with the effective input state. By computing the system compensation matrices from pullbacks taken over a period of eight months, we found that the system compensation matrices remained very stable in the clinical environment encountered in the catheterization laboratory and could be used interchangeably without dramatic deterioration of the reconstruction.

Another assumption of SIPS is the presence of sufficient PMD. Effectively, what matters is the spread on the Poincaré sphere of the input state probing the cumulative sample retardance, including the catheter-transmission. This spread is a combination of the amount of PMD and the relative orientation between the source state and the PMD eigenstate.

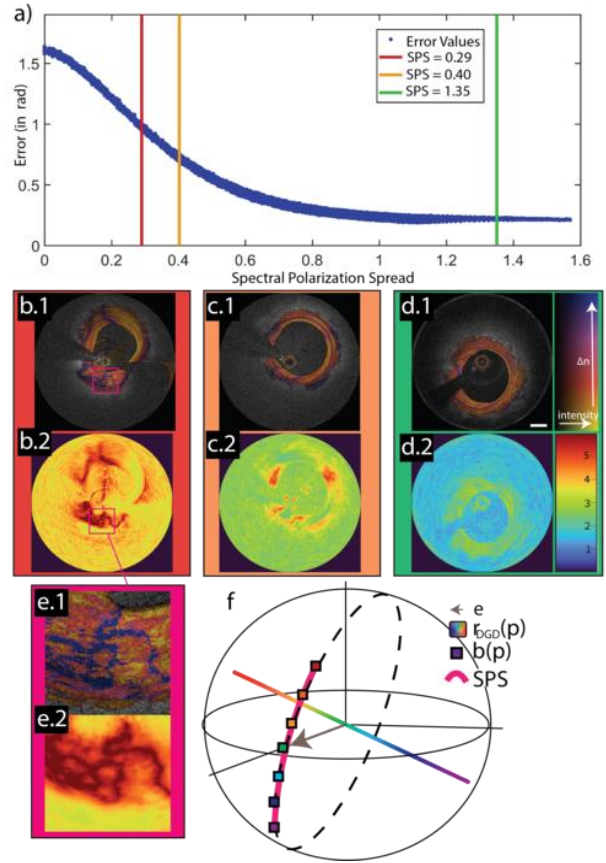

Figure 5: Exploration of the effect of system PMD-induced spectral spread on SIPS. a) Simulation results showing mean retardance error of SIPS-derived cumulative retardance as a function of spectral polarization spread (SPS). The red, orange, and green line represent the experimental SPS encountered in various real systems with a small (red), medium (orange) and large (green) SPS. The local birefringence images (b.1, c.1, d.1, e.1), and error metric (b.2, c.2, d.2, e.2) display how this effect changes the reliability of the SIPS metrics. Birefringence is scaled from 0 to $2.2 \cdot 10^{-3}$. The system with little spread (b &e) shows an increased error metric (b.2, e.2) and the local birefringence image exhibits significant artifacts (b.1, e.1). The system with intermediate spread still features some artifacts in the local birefringence image (c.1) although with an overall reduced error metric (c.2) compared to the system with less spread. Lastly, the system with sufficient spread, used throughout this study, shows little artifact (d.1) and a low overall error metric (d.2). f) Visualization of the spectral spread on the Poincare sphere. The input polarization state (brown arrow) is transformed by the system PMD (rainbow line) into a spectrally dependent polarization state b(p) (rainbow squares). The pink line indicates the SPS

The present work has been performed on lab-built systems with a suitable amount of PMD, which arises primarily from fiber circulators. Other systems may have smaller amounts of PMD. To investigate how much PMD is required, we estimated the error on cumulative retardance by simulating measurements, including an inherent signal to noise ratio of 10dB, varying system PMD and input state realizations, and using the maximum likelihood SIPS pipeline for reconstruction, shown in Fig. 5. The combined effect of the PMD magnitude and relative orientation to the input state was expressed as a measure of *spectral polarization spread,* visualized in Fig. 5.f), and defined as

$$\text{SPS} = \sum_p |\mathbf{b}_{p+1} - \mathbf{b}_p|, \qquad (26)$$

where $\mathbf{b}_p$ is the Stokes vector representing the input state of light after interacting with the system PMD

$$\mathbf{b}_p = \mathbf{Q}(p) \cdot \mathbf{e}. \qquad (27)$$

Fig. 5a) displays the average error across 5,000 simulated sample measurements for each of 100 random PMD orientations each with 100 PMD magnitudes. Each simulated measurement was for a random sample optic axis orientation and corresponding cumulative retardance that aligns the output state of the central bin with the mirror state of the system. There is a notable decrease in error as the SPS increases and thereby enhances the information extracted from the sample. The birefringence image of the system with reduced SPS has significant artifacts in localized areas. These artifacts are the result of an unfortunate combination of catheter transmission effects and local optical tissue properties, which results in an input state that is aligned with the optic axis of the cumulative round-trip retardance. This situation results in a poorly defined SVD decomposition matrix and high error on the subsequent estimated Jones matrix. To obtain a pixel-wise metric predicting the expected reconstruction error, we computed the condition number of the matrix $\Re\{\mathbf{H}\}$, which we define as the ratio of the largest singular value over the smallest singular value.

$$\text{Error Metric} = \log_{10}(\text{Condition Number}(\Re(\mathbf{H}))). \qquad (28)$$

The error metric is generally increased in systems with lower SPS (Fig. 5, b.2, c.2, d.2), but also locally highlights specific regions that are prone to artifacts, as clearly highlighted in Fig. 5e. We speculate that the error metric will serve to further refine retardance estimation in the future and suppress the resulting artifacts. At the least, the error metric will allow masking areas with unreliable birefringence measurements. All results presented for both the quantitative and qualitative analysis were acquired with a system with a spectral polarization spread of 1.35 with reduced artifacts compared to higher SPS systems.

An increased SPS will inherently degrade the point spread function of our system. However, the magnitude of the changes required to enable high-quality SIPS estimation is quite low compared to the resolution of OCT. The system with the greatest degradation observed analyzed in this paper has an overall SPS of 1.35, resulting in an illumination group delay of 1.37μm. This degradation of the point spread function is significantly smaller than the nominal axial resolution of our typical PS-OCT system (~9.4μm), so there is no significant loss of resolution.

In summary, our framework enables accurate retardance estimation with only a single input polarization state. SIPS capitalizes on the inherent wavelength-dependent effects of existing clinical systems to avoid the errors previously associated with single input measurements. We have developed an analytical, closed form solution to the maximum likelihood estimation of the sample retardance given the system parameters, which are directly obtained from typical measurements. Our framework makes an important step towards broadening the clinical availability of PS-OCT by enabling the extraction of polarimetric information, retrospectively or in future studies, from measurements performed with unaltered clinical systems.

## Acknowledgements

We thank Dr. Joost Daemen from the Erasmus Medical Center in Rotterdam for providing the intravascular PS-OCT data of the repeatability study, and Dr. Kenichiro Otsuka from the Osaka Metropolitan University Graduate School of Medicine for providing anonymized intravascular PS-OCT data acquired by systems with varying SPS.

## Data Availability

All reconstruction algorithms, sample processing scripts and test data are openly available in the CBORT-NCBIB/MLSIPS repository [31].

**Disclosures.** Massachusetts General Hospital and Nanyang Technological University have patent licensing arrangements with Terumo Corporation. Drs. Xiong, Liu, Bouma and Villiger have the right to receive royalties as part of the patent licensing arrangements with Terumo Corporation. Dr. Bouma has a financial interest in Soleron Imaging, LLC, a seller of unique optical imaging instruments and components used in this research. Dr. Bouma's interests were reviewed and are managed by Massachusetts General Hospital and Mass General Brigham in accordance with their conflict of interest policies.

## Supplemental Document

See Supplement 1 for supporting content.